\documentclass[aps,prc,preprintnumbers,twocolumn]{revtex4-1}

\usepackage{graphicx}

\begin{document}

\preprint{INR RAS 1301/2011}

\title{\bf Working characteristics of the New Low-Background Laboratory
(DULB-4900)
}

\author{
Ju.M.~Gavriljuk$^a$,
A.M.~Gangapshev$^{a}$, 
A.M.~Gezhaev$^{a}$,
V.V.~Kazalov$^{a}$, \\
V.V.~Kuzminov$^{a}$,
S.I.~Panasenko$^{b}$,
S.S.~Ratkevich$^{b}$,
S.P.~Yakimenko$^{a}$
\\
$^a$ \small{\em Baksan Neutrino Observatory INR RAS} \\
$^b$ \small{\em Karazin Kharkiv National University, Ukraine} \\
}

\begin{abstract}
A concise technical characteristic of a new low-background laboratory DULB-4900 of the BNO INR RAS is presented. The technique and the results of background measurements in the Hall, ordinary box and low-background box are presented. $^{222}$Rn contamination in the laboratory air has been measured by direct detection of  $\gamma$-radiation of its daughter $^{214}$Bi distributed over the volume of the low-background box. The results of the data analysis are presented.
\end{abstract}

\maketitle


\section{Introduction}

Several low-background experiments  are carried  out in the Baksan Neutrino Observatory INR RAS, searching for rare processes of double beta decay of $^{136}$Xe \cite{r1}, for 2K-capture for $^{78}$Kr and $^{124}$Xe \cite{r2} and \cite{r3}. The best half-life limits have been obtained in these experiments thereby closing a number of theoretical models. Low-background shielding is one of the basic elements of low-background facilities which significantly suppresses the background produced by $\gamma$-quanta and neutrons from natural radioactive contamination coming from the surrounding rocks and construction materials. It also provides the isolation of the detector from the radioactive $^{222}$Rn which is always present in the environmental air.  A specific shielding has been designed and constructed for each detector for the experiments mentioned above. The total mass of each shielding is about  $30\div40$ tons.  The working volume of the detectors is about several liters with a mass of gaseous radioactive isotope samples from  $\sim0.1$ to $\sim1$ $kg$. Sensitivity of detectors with working volume of $\sim10$ l is not high enough  to continue experiments, and new low-background gaseous detectors of $1\div3$ $m^3$ volume allowing one to study samples of $10\div50$ $kg$ mass have been proposed. It is with these new detectors that some of the above mentioned rare processes could be discovered. In order to locate such detectors, the inner fiducial volume of the shielding should be about 20 $m^3$.

\section{The description new low-background laboratories}

A new and unique deep underground low-background laboratory (DULB-4900) has come into operation in the BNO INR RAS. The laboratory is located at a distance of 3700 $m$ from the main entry to the Baksan Neutrino Observatory tunnel, in the hall of $\sim6\times6\times40$ $m^3$. Thickness of the mountain rock over DULB corresponds to 4900 $m.w.e.$, thereby decreasing cosmic ray flux by $\sim10^7$ times \cite{r4}.
There are two adjacent low-background boxes (l.b. boxes) and six ordinary boxes. Ordinary boxes are with walls of sheet steel.
The walls, doors, floors and ceilings of the two low-background boxes are composed of 250 $mm$ polyethylene $+1$ $mm$ Cd $+150$ $mm$ high-purity Pb with the total mass of the shielding material $30 t$, $1 t$ and $190 t$, respectively. The outside and inside surfaces of the l.b. boxes are lined up with sheet steel of 3 $mm$ thickness. To prevent the steel from corrosion the inside surface of steel sheets was painted with colorless lacquer of low radioactive contamination level. The dimensions of the inner shielded volume of each of l.b. boxes is $2.5\times3.2\times2.5$ $m^3$. In the working space of the l.b. boxes, the flux of $\gamma$-quanta and neutrons is reduced by hundreds of times. Such significant suppression, in these two boxes, of cosmic ray flux and radiation of natural radioactive elements from the environmental rock and air (the basic source of background of the ionizing - radiation detectors) opens new perspectives to search for extremely rare nuclear processes using massive targets.

The radioactive environment in the l.b. boxes allows one to search for candidates (particles) for the 'dark' mass of the Universe using various types of detectors: gas ionizing detectors with volume of $\sim2$ $m^3$, solid-state scintillation detectors with mass of $\sim300$ $kg$, two-phase emission detector based on liquified xenon with mass of $\sim100$ $kg$.

The same detectors could be used in search for neutrinoless double beta-decay of $^{136}$Xe isotopes (a gas-filled or two-phase detector) and $^{100}$Mo (scintillation detector with $^{40}$Ca$^{100}$MoO$_4$ crystals), both having half-life sensitivity of about $10^{26}$ years (5 years of measurements).

Investigations of various effects produced in the living organisms in the low-level radioactive environment, study of natural radioactive element distribution in the human body are some challenges offered in these laboratories for biologists.

The laboratory is well equipped with everything necessary for its normal functioning: electricity, cold water supply, sewer system, telephone line, optical coupling, forced ventilation, lifting mechanisms, local railway, auxiliary rooms, and testing facilities.

\begin{figure}[pt]
\includegraphics*[width=2.5in,angle=0.]{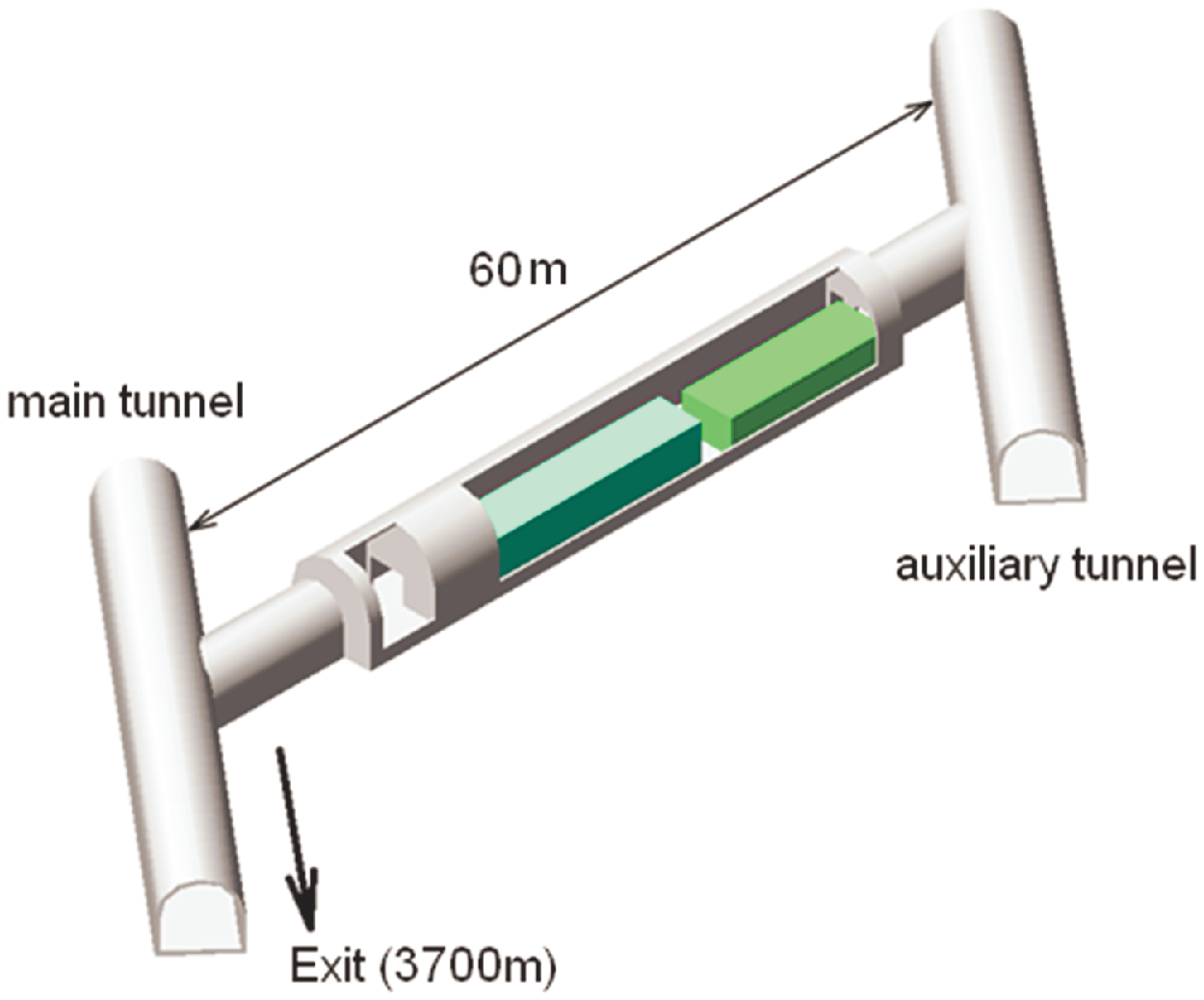}%
\caption{\label{p00} DULB and tunnel lay-out.}
\end{figure}

\begin{figure}[ht]
\begin{center}
\includegraphics*[width=3.5in]{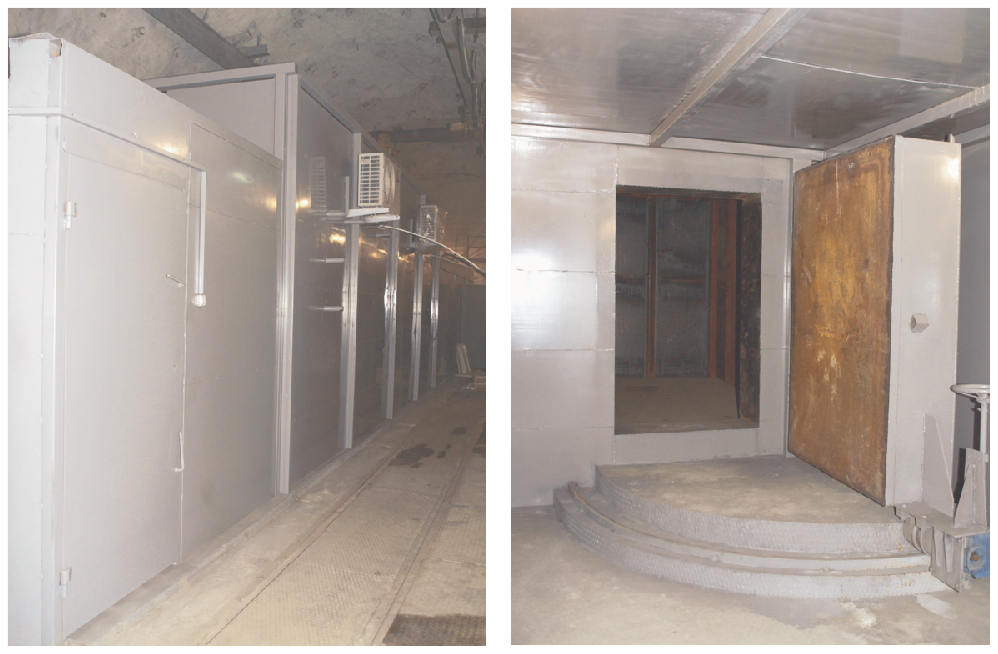}%
\caption{\label{f1} General view of auxiliary and low-background boxes.}
\caption{\label{f2} Interior view of the low-background box with a door opened.}
\end{center}
\end{figure}

Schematic longitudinal section of the laboratory is presented in Fig.\ref{p00}.
General view is given in Fig.\ref{f1} and Fig.\ref{f2}.


\section{The technique of measurement and $\gamma$-background obtained in the low-background boxes}

Measurement of $\gamma$-radiation in the l.b. boxes has been performed using the scintillation detector based on NaI(Tl) crystal (9.72 $kg$, d=150 $mm$, h=150 $mm$) encapsulated by stainless steel sheets. The crystal is viewed with photomultiplier PMT-49 mounted on the quartz output window.

\begin{figure}[pt]
\includegraphics*[width=2.25in,angle=270.]{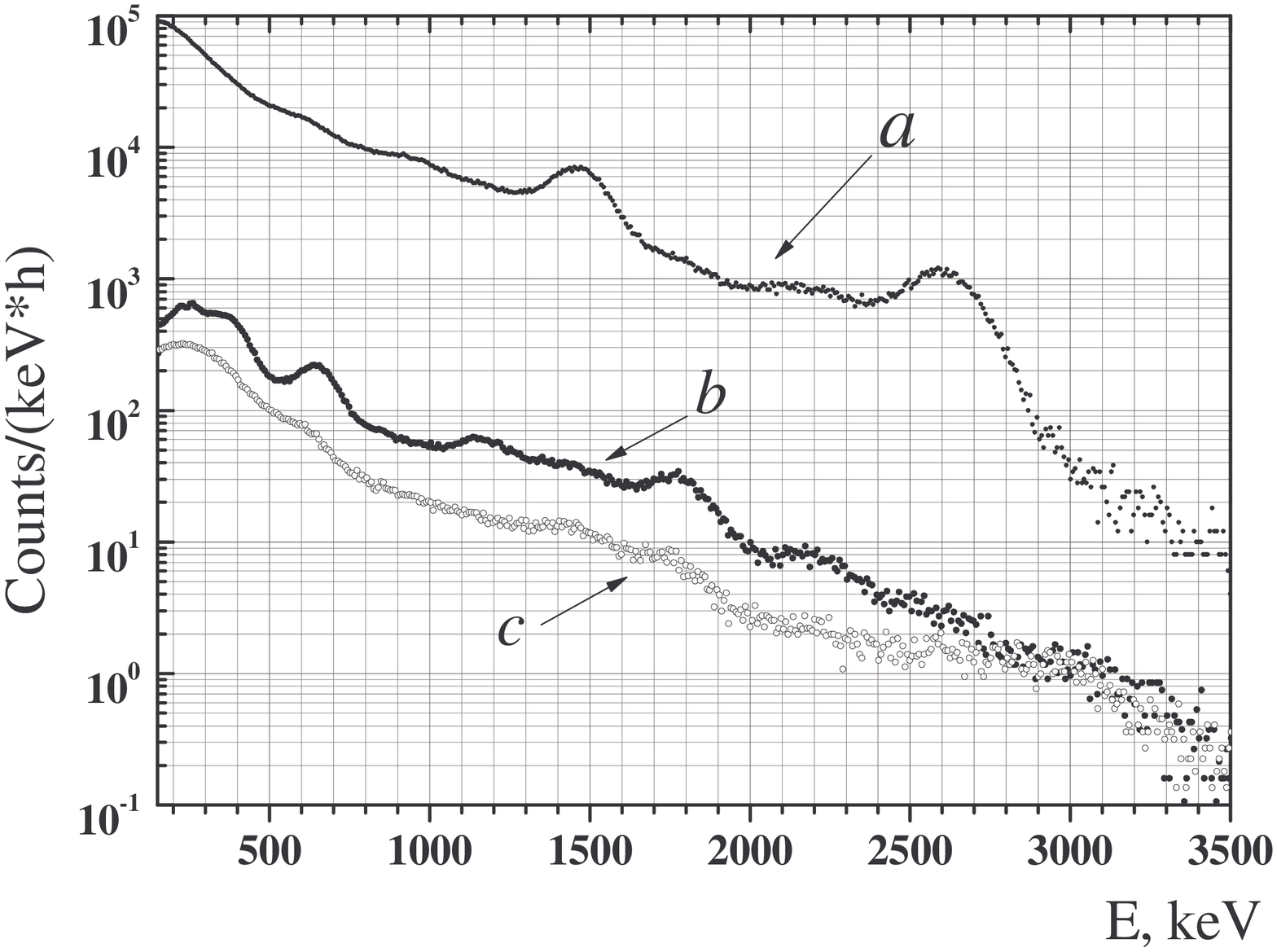}%
\caption{\label{p4} Background spectra in the unshielded room of DULB (\emph{a}), in a low-background box (\emph{b}), in a low-background box with additional lead shielding(\emph{c}).}
\end{figure}

In Fig.\ref{p4}, spectrum $"\emph{c}"$ demonstrates the intrinsic background of the detector obtained during
22.3 hours and reduced to 1 hour.
The background of the detector, tightly surrounded  by additional shielding of lead of 150 $mm$ thickness, measured in one of the l.b. boxes was taken to be the intrinsic background of the detector. The additional shielding has been constructed to screen $\gamma$-radiation generated by $^{214}$Bi isotope which is the daughter product of  $^{222}$Rn.
The measurements \cite{r5} give the activity of $^{40}$K, $^{232}$Th and $^{238}$U in the lead has been evaluated as $\leq5.2\times10^{-3}$, $\leq1.4\times10^{-3}$, and $(1.2\pm0.2)\times10^{-2}$ Bq/$kg$, respectively. In this lead, there is a detectable admixture  of radioactive isotope of $^{210}$Pb (T$_{1/2} = 22.3$ years, $\beta^-$ , E$_{\beta max}=61$ keV) $\rightarrow$ $^{210}$Bi (T$_{1/2}=5.01$ days, $\beta^-$, E$_{\beta max}=1160$ keV) $\rightarrow$ $^{210}$Po (T$_{1/2}=138.4$ days, $\alpha$) $\rightarrow$ $^{206}$Pb. Bremsstrahlung effect due to electrons of daughter $^{210}$Bi  decay adds to the background of the detector at energies lower than  $\sim1000$ keV. Counting rate in the energy range of $0.3\div3.0$ MeV was 2.9 s$^{-1}$. One can see lines of 1460 keV($^{40}$K ) and 1765 keV ($^{214}$Bi) in spectrum $"\emph{c}"$ (Fig.\ref{p4}).

\begin{figure}
\includegraphics*[width=2.25in,angle=270.]{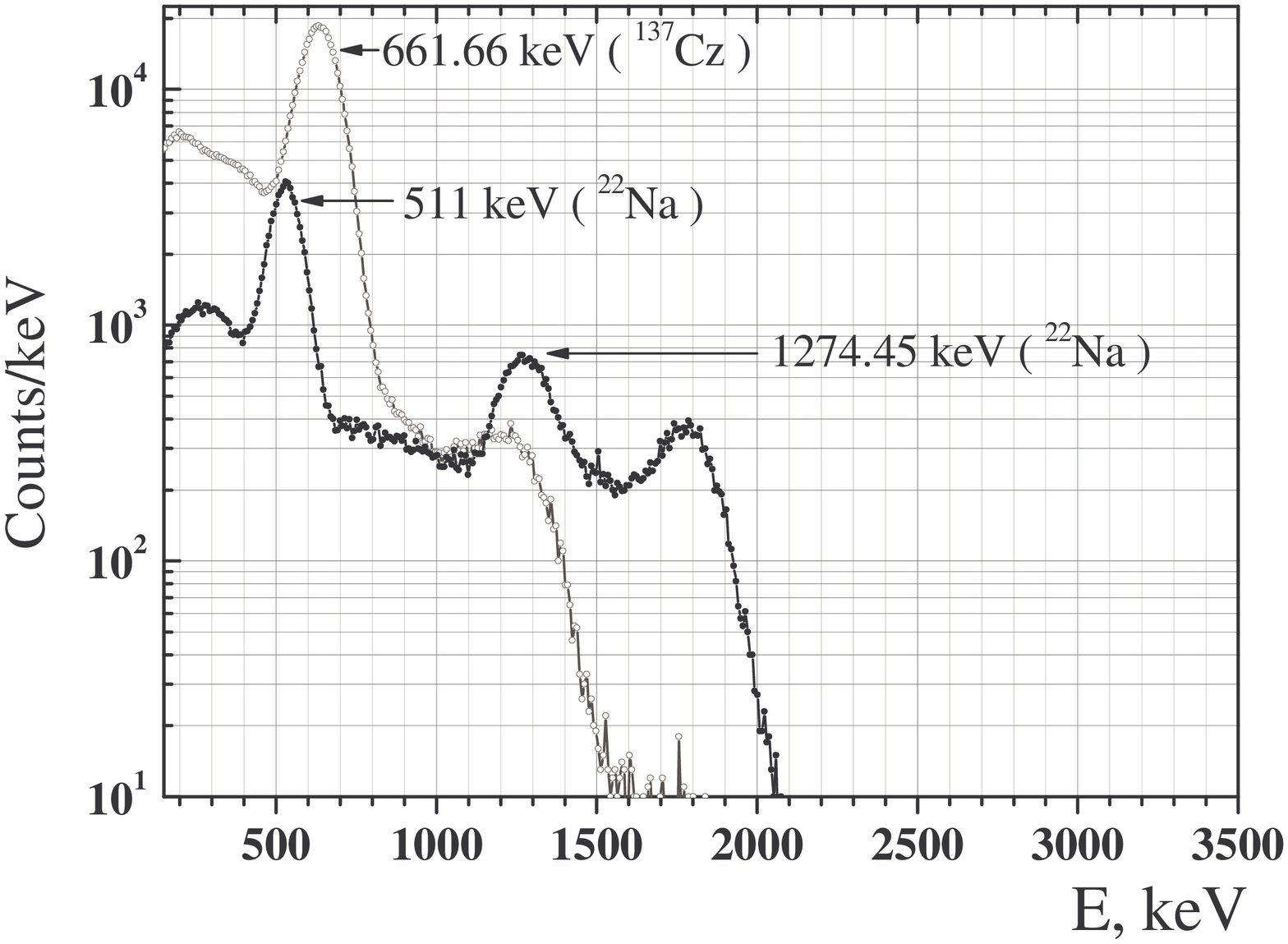}%
\caption{\label{p5} Spectra of $^{22}$Na and $^{137}$Cs sources.}
\end{figure}

The $\gamma$-quanta background of $^{40}$K is mainly due to the presence of this isotope in the glass of PMT-49. This background component could
be diminished by placing an additional light guide of sufficient length between the PMT and the crystal. The $\gamma$-quanta
flux coming to the crystal would be decreased with the decrease of geometrical factor and absorption of radiation in the
optical fiber material.  The $^{214}$Bi lines, present in the background spectra, could be explained by  the PMT construction materials,
preamplifier parts, lead shielding and radon of the air left in the residual cavities in the lead shielding.
Some of $^{214}$Bi atoms due to the radon decay in the air are also produced on the surface of the casing encapsulating the crystal and on the inside walls of the shielding. The intermediate daughter atoms of $^{218}$Po generated in the ionized form could deposit on the grounded surfaces where $^{214}$Bi is finally produced in the decay chain $^{222}$Rn$\rightarrow^{218}$Po$\rightarrow^{214}$Pb$\rightarrow^{214}$Bi. High contrast of 609 keV-line could be interpreted
as a sign that all  $^{214}$Bi  has decayed in the air within cavities. The absence of intermediate absorbing
material between the source and the crystal makes the intensities and shapes of the lines of  $^{214}$Bi  to look like lines
caused by some close calibration source. As an example, Fig.\ref{p5} presents the spectra of $^{137}$Cs   and $^{22}$Na calibration sources measured
using the shielded detector.

When a source is distributed over the volume of the substance that has a large absorption coefficient of gamma radiation (e.g. lead), the ratio of intensities of the lines of different energies emitted by the source would differ from the case above.  The difference is due to:  1) complexity of the dependence of gamma-ray absorption cross-section on quantum energy and atomic number of the absorber material;  2) increase in radiation absorption with decrease in energy of radiation; 3) abundance, in the spectra, of quanta, of lesser, than that of the base line,energy appearing in Compton scattering of the base lines in the material.

In the crystal, the relative content of $^{214}$Bi of the  $^{238}$U chain is small. This conclusion has been made after the analysis of the contribution to the detector background of $\alpha$-particles with energies of $5.0\div7.7$ MeV due to the decays of the elements of the Uranium chain. The amplitude of the signal of these $\alpha$-particles is expected to be registered in the range of $3.5\div5.4$ MeV (in terms of electron energy scale), because $\alpha/e$-ratio for NaI(Tl)  is $\sim0.7$.

In order to determine the suppression coefficient of the $\gamma$-ray background in the l.b. boxes one has to know the level of the background
in the boxes without shielding. To make the necessary measurement the detector was taken out of the measuring set-up and placed in the separate
light-intercepting and screening cylindrical casing  made of stainless
steel of 0.7 $mm$ thickness. This assembly was placed on the floor of
the ordinary box with unshielded walls in the l.b. box.
The measurement was taken during 0.5 h. The spectrum thus obtained and reduced to one hour is spectrum "\emph{a}" in Fig.\ref{p4}. Counting rate in the energy interval of $0.3\div3.0$ MeV is 695.8 s$^{-1}$. Main peaks in the spectrum correspond to the lines of 1460 keV ($^{40}$K) and 2614 keV ($^{208}$Tl).

In order to measure the $\gamma$-radiation background in the l.b. boxes the assembly has been suspended in the middle of the inner volume of the l.b. box. There is no extraneous materials in the assembly expect for the attached to the interior wall cooling board
There is no extraneous materials in the assembly expect for the attached to the interior wall cooling board of the air-conditioner
of the air-conditioner.
The measurement was taken during 18.8 h. Counting rate in the energy region of $0.3\div3.0$ MeV was 6.23 s$^{-1}$. The spectrum reduced to one hour is spectrum $"\emph{b}"$ in Fig.\ref{p4}.  Main peaks in this spectrum correspond to $\gamma$-lines of $^{214}$Bi  source. Their shape corroborates the fact argumented above, that the source is either in the air or on the interior walls of the box, and is due to $^{222}$Rn decay in the air inside the box. Though some amount of $^{214}$Bi could penetrate with the flow of the outdoor air  through the specific holes in the walls of the box. In the case of placing the ultra-low-background device in the l.b. box to search for extremely rare decay processes one has to provide  special purification of the inside air from radon contamination.

Table \ref{tab1}   presents    measured counting rates of the detector for different energy ranges used to obtain numerical data for the
$\gamma$-ray background
suppression coefficients.

\begin{table}
\caption{\label{tab1} Detector counting rate for all measurements taken.}
\begin{tabular}{l|c|c|c|c}
  \hline
  Energy, MeV           & 0.3 -1.0 & 1.0 - 1.5 &1.5 - 2.0 & 2.0 - 3.0 \\
  \hline
  \hline
  Counting rate,          s$^{-1}$ \\
  \hline
  Apparatus compartment & 623.0    & 133.3 & 60.5   & 34.6    \\
  l.b. box              & 4.91     & 0.841 & 0.364  & 0.116   \\
  l.b. box + Pb         & 2.45     & 0.329 & 0.122  & 0.062   \\
  \hline
\end{tabular}
\end{table}

The ratio of the counting rate in the ordinary box to the intrinsic background of the detector for presented in the Table I energy regions is 254; 404; 496; 558, respectively. The same ratio for the background in the low-background box is 127; 159; 166; 298, respectively.

\section{The technique of measurement and the results obtained for the content of Rn in the air of DULB-4900}

A low-background gamma-ray detector placed in the center of the shielding with a large inner volume ( $\sim20$ m$^2$) is an experimental set-up to continuously measure the radon activity in the environmental air by detecting gamma-radiation from its daughter short half-life isotopes ($^{21}$Bi). The total amount of radon (and $^{214}$Bi) is proportional to the volume of the box, and counting rate of the detector is proportional to the mean distance from the detector to the wall of the chamber. It is easy to understand the latter statement when you imagine the detector  in the center of the facility surrounded by concentric layers of air with equal thickness. The volume of air and quantity of radioactive atoms in a layer increase in an approximately proportional way to the surface of the layer (or to the square of the layer's radius). The solid angle through which the detector could be viewed from each point of the layer is in inverse proportion to the square of the layer's radius. Therefore, the number of $\gamma$-quanta, coming into the detector from each layer, remains constant and their total number is in proportion to the number of layers or their sum thickness (mean transit radius). Radiation absorption in the air is negligible.

High sensitivity of such a detector and the described technique allows one to obtain the best marginal results.

\begin{figure}[pt]
\includegraphics*[width=2.25in,angle=270.]{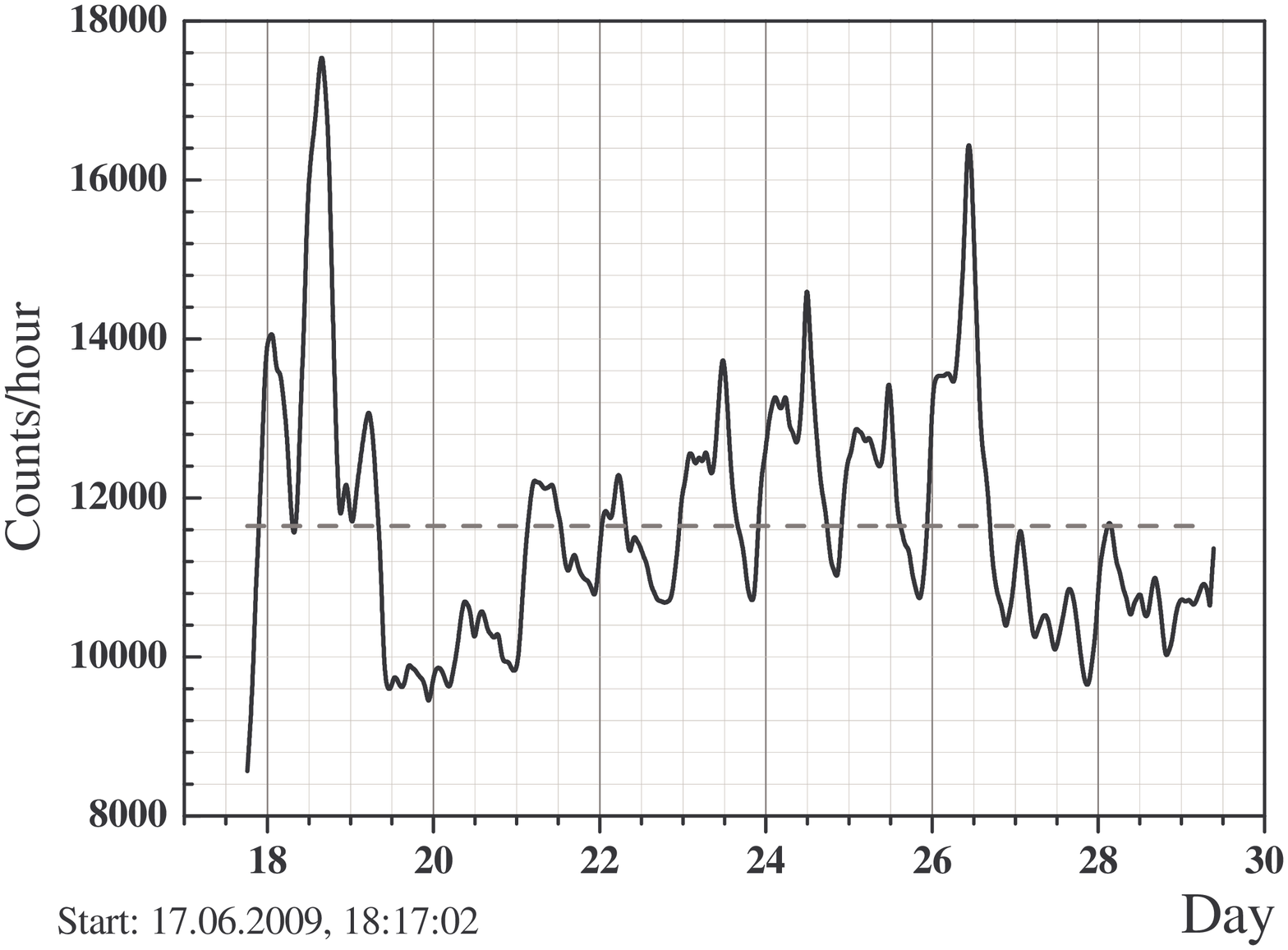}%
\caption{\label{p6} Dependency on time of the counting rate of the detector suspended in the center of the ventilated l.b. box.}
\end{figure}

To test the sensitivity of the facility to radon variations in the air by the detection of  $\gamma$-radiation due to $^{214}$Bi
one of the specific holes was equipped with a fan blower continuously fanning the air from the main hall of the DULB
into the box ($\sim1$ m$^3\times$min$^{-1}$) during 12 days. Pulses from the detector (suspended in the center of the l.b. box)
were registered by a digital oscilloscope LA-n20-12PCI incorporated within a personal computer.
A spectrum collected during one hour was recorded in an individual file. In processing the data the counts in an individual
spectrum were summed up above the threshold of 482 keV. The sequence of data thus obtained is presented in Fig.\ref{p6}.
Mean counting rate above the threshold was 11648 h$^{-1}$ for the total period of measurements.
The corresponding activity of $^{222}$Rn was calculated in view of its equilibrium with activity of daughter $^{214}$Bi distributed
uniformly in the air of the l.b. box. $^{214}$Bi activity was evaluated using line 609 keV. The registration efficiency of 609 keV-radiation
calculated with GEANT 4 is 0.12\%, in a photopeak. The averaged counting rate under the photopeak with background subtracted
was found to be 1404 h$^{-1}$. Given the yield of  $\gamma$-quanta of 609 keV per one decay of $^{214}$Bi (0.461) one can estimate
the volumetric activity for $^{214}$Bi and consequently for $^{222}$Rn:  $(1404/3600/0.0012/0.461/20)=(35.2\pm1.3)$ Bq$\times$m$^{-3}$.
The intrinsic background of the detector above the threshold of 482 keV is 5120 h$^{-1}$.
The average input of $^{214}$Bi decays into the total background of the detector in the l.b. box is 6528 h$^{-1}$. Variations of
$^{214}$Bi  activity (i.e. $^{222}$Rn content) for the whole period of measurements are within $^{+92}_{-33}$\% of the mean value.
Among different factors of radon variation in the air of the main hall, the basic one is considered to be the ventilation of the main tunnel (forced ventilation by the outdoor air due to the external exhaust blower in the auxiliary tunnel) and the main hall (air flow due to the difference in the air pressure of the main tunnel and the auxiliary tunnel and variable size of the tunnel passage sections).  During the whole period of measurements (17.06.-29.06.2009), there happened to be a complete stoppage of work of the external exhaust blower (18.06.2009), and as a result the flowing of the air with higher radon content from the auxiliary tunnel into the main hall occurred. During the working hours, the main doors of the DULB-4900 laboratory remained open, while at night and on weekdays they were closed and as a consequence the air flow rate decreased. In addition to that, the opening and closing of the lock-gates between the two tunnels, when the shift arrived or some operations were carried out in the tunnel, lead to 'jumps' in the air pressure of the main tunnel.  At present, it does not seem feasible to take into account all the factors causing the radon variation in the air.
One can see in Fig.\ref{p6} the significant decrease of radon variation on weekends (20-21.06., 27-28.06.).

\section{Main results and conclusions}
\begin{enumerate}
    \item Background measurements for the scintillation detector with crystal NaI(Tl) (d = 150 $mm$, h = 150 $mm$) have been carried out in different environmental conditions in Deep Underground Low-Background laboratory (DULB) of the Baksan Neutrino Observatory INR RAS.
    \item Background spectra for the detector have been obtained in the unshielded box, in the l.b. box, and in the l.b. box with additional 15 cm lead shielding.  In the latter case calibration measurements have been carried out with $^{22}$Na and $^{137}$Cs sources.
    \item Suppression coefficients for background of the tested  low-background detector with shieldings of different composition have
    been obtained. Maximal background suppression in the l.b. box with additional shielding is $\sim600$ for the range of 2.0-3.0 MeV.
    The average suppression coefficient is  287 for the energy range of 0.3-3.0 MeV. Relatively low suppression coefficient
    of the background could be explained by presence of radioactive admixtures in the detector and test facility.
    \item Background of the detector in the l.b. box is mainly due to $^{222}$Rn and its daughter  $^{214}$Bi  in the the air.
    In the energy range of 0.3-2.0 MeV the input of the air radioactive contamination into the background of the detector for
    the considered in this paper conditions of measurements is $\sim56$\% of the mean counting rate.
\end{enumerate}

The testing facility, consisting of the l.b. box and detector suspended in its center, allows one to probe the marginal, as to its sensitivity and efficiency, possibilities of the described in this paper technique to monitor the  radon content in the environmental air by registering $\gamma$-radiation of the daughter $^{214}$Bi.

\bm{Acknowledgments}
The authors gratefully acknowledge the cooperation of the whole stuff of the Low-background laboratory of the BNO INR RAS in the measurements taken.
This work was partially supported by the Federal goal-oriented program "Research and development of the priority lines of the scientific and technical complex of Russia for the 2007-2013 years" of the Russian Ministry of Education and Science (contract No 16.518.11.7072).

\end{document}